\documentclass[10pt,conference, two column]{IEEEtran}
\IEEEoverridecommandlockouts
\usepackage[left=0.7in, right=0.84in, bottom=1in, top=0.7in]{geometry}

\makeatletter
\newcommand{\AddInputPath}[1]{%
  \ifx\input@path\@undefined
    \def\input@path{#1}
  \else
    \g@addto@macro{\input@path}{#1}
  \fi
}
\makeatother
\AddInputPath{{../}}

\usepackage{url}
\usepackage{shellesc}

\usepackage{balance}
\usepackage{etex}

\usepackage{relsize}

\usepackage[dvipsnames,svgnames,usenames]{xcolor}
\usepackage{array}
\usepackage{booktabs,tabularx}
\usepackage{multirow}
\usepackage[per-mode=symbol,detect-mode=true]{siunitx}
\usepackage{graphicx}

\usepackage[T1]{fontenc}
\usepackage{textcomp}
\usepackage[utf8]{inputenc}
\usepackage[final]{microtype}
\usepackage{icomma}
\usepackage{xspace}

\usepackage[tbtags]{amsmath}
\usepackage{amssymb,amsfonts,bm}
\usepackage{mathtools} 
\usepackage{dsfont}
\usepackage{mathrsfs}
\usepackage{accents}
\usepackage{empheq}
\usepackage{nccmath}

\usepackage{setspace}
\usepackage{bbm}

\usepackage{color}
\usepackage{calc}
\usepackage{tikz}
\usepackage{pgfplots,pgfplotstable}
\usetikzlibrary{pgfplots.groupplots}
\usetikzlibrary{matrix} 
\pgfplotsset{compat=1.18} 
\usepackage{pdftexcmds}
\makeatletter
\newcommand{\strequal}[2]{\pdf@strcmp{#1}{#2}==0}
\makeatother

\usepackage[capitalize]{cleveref}
\usepackage[font=small]{subcaption}

\usepackage[inline]{enumitem}
\usepackage{algorithm}
\usepackage{algpseudocode}
\makeatletter
\newcommand{\algmargin}{\the\ALG@thistlm}
\makeatother
\newlength{\whilewidth}
\algdef{SE}[parWHILE]{parWhile}{EndparWhile}[1]
  {\parbox[t]{\dimexpr\linewidth-\algmargin}{%
     \hangindent\whilewidth\strut\algorithmicwhile\ #1\ \algorithmicdo\strut}}{\algorithmicend\ \algorithmicwhile}%
\algnewcommand{\parState}[1]{\State%
  \parbox[t]{\dimexpr\linewidth-\algmargin}{\strut #1\strut}}

\usepackage{glossaries}
\usepackage{ifthen}
\usepackage[noadjust]{cite}
\usepackage{multibib}

\usepackage{comment}
\usepackage{todonotes}
\let\legacytodo\todo
\newcommand{\ruggedtodo}[2][]{\tikzexternaldisable\legacytodo[#1]{#2}\tikzexternalenable}
\renewcommand{\todo}[1]{\ruggedtodo[inline]{#1}}
\usetikzlibrary {arrows.meta}

\makeglossaries

\newacronym{fso}{FSO}{free-space optical communication}
\newacronym{tle}{TLE}{two-line element}
\newacronym{ai}{AI}{artificial intelligence}
\newacronym{ann}{ANN}{artificial neural network}
\newacronym{jscc}{JSCC}{joint source-channel coding}
\newacronym{raan}{RAAN}{right ascension of the ascending node}
\newacronym{uav}{UAV}{unmanned aerial vehicle}
\newacronym{haps}{HAPS}{high-altitude platform station}
\newacronym{6g}{6G}{sixth generation}
\newacronym{cgr}{CGR}{contact graph routing}
\newacronym{dtn}{DTN}{delay-tolerant networking}
\newacronym{fl}{FL}{federated learning}
\newacronym{fo}{FO}{federated optimization}
\newacronym{dl}{DL}{deep learning}
\newacronym{fedavg}{FedAvg}{federated averaging}
\newacronym{dml}{DML}{distributed ML}
\newacronym{ps}{PS}{parameter server}
\newacronym{ml}{ML}{machine learning}
\newacronym{sgd}{SGD}{stochastic gradient descent}
\newacronym{dsgd}{DSGD}{distributed stochastic gradient descent}
\newacronym{isl}{ISL}{inter-satellite link}
\newacronym{gsl}{GSL}{ground-satellite link}
\newacronym{gs}{GS}{ground station}
\newacronym{ecef}{ECEF}{earth-centered, earth-fixed}
\newacronym{eci}{ECI}{Earth-centered inertial}
\newacronym{ofdm}{OFDM}{orthogonal frequency-division multiplexing}
\newacronym{los}{LOS}{line-of-sight}
\newacronym{leo}{LEO}{low Earth orbit}
\newacronym{meo}{MEO}{medium Earth orbit}
\newacronym{gso}{GSO}{geosynchronous orbit}
\newacronym{geo}{GEO}{geostationary}
\newacronym{ntn}{NTN}{non-terrestrial networks}
\newacronym{eo}{EO}{Earth observation}
\newacronym{iot}{IoT}{Internet of Things}
\newacronym{irs}{IRS}{intelligent reflecting surface}
\newacronym{socp}{SOCP}{second-order cone program}
\newacronym{soc}{SOC}{second-order cone}
\newacronym{dsl}{DSL}{digital subscriber line}
\newacronym{wsee}{WSEE}{weighted sum energy efficiency}
\newacronym{mmwave}{mmWave}{millimeter wave}
\newacronym{dfg}{DFG}{Deutsche Forschungsgemeinschaft}
\newacronym{haec}{HAEC}{Highly Adaptive Energy-Efficient Computing}
\newacronym{hpc}{HPC}{High Performance Computing}
\newacronym{mac}{MAC}{multiple-access channel}
\newacronym{bc}{BC}{broadcast channel}
\newacronym{siso}{SISO}{single-input single-output}
\newacronym{simo}{SIMO}{single-input multiple-output}
\newacronym{miso}{MISO}{multiple-input single-output}
\newacronym{mimo}{MIMO}{multiple-input multiple-output}
\newacronym{af}{AF}{amplify-and-forward}
\newacronym{df}{DF}{decode-and-forward}
\newacronym{cf}{CF}{compress-and-forward}
\newacronym{mwrc}{MWRC}{multi-way relay channel}
\newacronym{dmmwrc}{DM-MWRC}{discrete memoryless multi-way relay channel}
\newacronym{pde}{PDE}{partial data exchange}
\newacronym{fde}{FDE}{full data exchange}
\newacronym{iid}{i.i.d.\@}{independent and identically distributed}
\newacronym{di}{DI} {difference of increasing}
\newacronym{dc}{DC}{difference of convex functions}
\newacronym{mm}{MM}{mixed monotonic}
\newacronym{mmp}{MMP}{mixed monotonic programming}
\newacronym{awgn}{AWGN}{additive white Gaussian noise}
\newacronym{wgn}{WGN}{white Gaussian noise}
\newacronym{awg}{AWG}{additive white Gaussian}
\newacronym{sic}{SIC}{successive interference cancellation}
\newacronym{snr}{SNR}{signal-to-noise ratio}
\newacronym{sinr}{SINR}{signal to interference plus noise ratio}
\newacronym{inr}{INR}{interference to noise ratio}
\newacronym{zf}{ZF}{zero-forcing}
\newacronym{mrt}{MRT}{maximum ratio transmission}
\newacronym{mmse}{MMSE}{minimum mean square error}
\newacronym{sud}{SUD}{single user decoding}
\newacronym{dof}{DoF}{degrees of freedom}
\newacronym{gdof}{GDoF}{generalized degrees of freedom}
\newacronym{nnc}{NNC}{noisy network coding}
\newacronym{dmn}{DMN}{discrete memoryless network}
\newacronym{csi}{CSI}{channel state information}
\newacronym{pmf}{pmf}{probability mass function}
\newacronym{dmic}{DM-IC}{discrete memoryless interference channel}
\newacronym{ic}{IC}{interference channel}
\newacronym{gic}{GIC}{Gaussian interference channel}
\newacronym{if}{IF}{interference}
\newacronym{ee}{EE}{energy efficiency}
\newacronym{gee}{GEE}{global energy efficiency}
\newacronym{tin}{TIN}{treating interference as noise}
\newacronym{snd}{SND}{simultaneous non-unique decoding}
\newacronym{sd}{SD}{simultaneous decoding}
\newacronym{hk}{HK}{Han-Kobayashi}
\newacronym{rs}{RS}{rate splitting}
\newacronym{rf}{RF}{radio frequency}
\newacronym{pa}{PA}{power amplifier}
\newacronym{lna}{LNA}{low noise amplifier}
\newacronym{lo}{LO}{local oscillator}
\newacronym{adc}{ADC}{analog-to-digital converter}
\newacronym{dac}{DAC}{digital-to-analog converter}
\newacronym{dsp}{DSP}{digital signal processing}
\newacronym{brd}{BRD}{best response dynamics}
\newacronym{br}{BR}{best response}
\newacronym{ne}{NE}{Nash equilibrium}
\newacronym{lhs}{LHS}{left-hand side}
\newacronym{rhs}{RHS}{right-hand side}
\newacronym{ran}{RAN}{radio access network}
\newacronym{qos}{QoS}{Quality of Service}
\newacronym{ngmn}{NGMN}{Next Generation Mobile Networks}
\newacronym{cap}{CAP}{Capacity Adaptation}
\newacronym{bwa}{BW}{Bandwidth Adaptation}
\newacronym{prb}{PRB}{physical resource block}
\newacronym{se}{SE}{spectral efficiency}
\newacronym{tp}{TP}{throughput}
\newacronym{bs}{BS}{base station}
\newacronym{ue}{UE}{user equipment}
\newacronym{mop}{MOP}{multi-objective optimization problem}
\newacronym{gda}{GDA}{generalized Dinkelbach's algorithm}
\newacronym{midcp}{MIDCP}{mixed integer disciplined convex programming}
\newacronym{lp}{LP}{linear program}
\newacronym{brb}{BRB}{branch reduce and bound}
\newacronym{bb}{BB}{branch and bound}
\newacronym{sit}{SIT}{successive incumbent transcending}
\newacronym{oma}{OMA}{orthogonal multiple access}
\newacronym{noma}{NOMA}{non-orthogonal multiple access}
\newacronym{wlog}{w.l.o.g.\@}{without loss of generality}
\newacronym{lsc}{l.s.c.\@}{lower semi-continuous}
\newacronym{usc}{u.s.c.\@}{upper semi-continuous}
\newacronym{kkt}{KKT}{Karush-Kuhn-Tucker}
\newacronym{ptp}{PTP}{point-to-point}
\newacronym{fspl}{FSPL}{Free space path loss}
\newacronym{sfl}{SFL}{satellite federated learning}
\newacronym{gu}{GU}{global update}
\newacronym{cu}{CU}{cluster update}
\newacronym{dod}{DoD}{depth of discharge}
\newacronym{sca}{SCA}{successive convex approximation}

\glsenableentrycount
\makeglossaries
\usetikzlibrary{positioning}
\usetikzlibrary{calc}
\usetikzlibrary{math}
\usetikzlibrary{fit}
\usetikzlibrary{intersections}
\usetikzlibrary{decorations.pathreplacing}
\usetikzlibrary{decorations.markings}
\usetikzlibrary{3d,angles}
\usetikzlibrary{arrows.meta}

\pgfdeclarelayer{background}
\pgfsetlayers{background,main}

\usetikzlibrary{external}

\usepgfplotslibrary{colorbrewer}

\tikzset{
	small1/.style={fill=DeepPink},
	small2/.style={fill=DeepSkyBlue},
	small3/.style={fill=MediumSpringGreen},
	ps/.style={fill=Gold},
	link/.style = {semithick},
	plane/.style={plane origin={(#1,0,0)}, plane x = {(#1,0,1)}, plane y = {(#1,1,0)}, rotate around y = -9, canvas is plane}
}

\tikzset{
	antenna/.pic={
		\draw[thick] (0,0) -- ++(120:2mm) -- ++(0:2mm) -- cycle -- (0,-1.5mm);
	}
}


\crefname{equation}{}{}
\crefrangeformat{equation}{(#3#1#4)--(#5#2#6)}
\crefmultiformat{equation}{(#2#1#3)}{ and~(#2#1#3)}{, (#2#1#3)}{, (#2#1#3)}
\crefrangemultiformat{equation}{#3(#1)#4--#5(#2)#6}{, #3(#1)#4--#5(#2)#6}{, #3(#1)#4--#5(#2)#6}{, #3(#1)#4--#5(#2)#6}

\DeclareMathOperator*\argmin{arg\,min}

\undef\mod
\DeclareMathOperator\mod{mod}

\let\vec\bm

\allowdisplaybreaks[3]

\DeclareSIUnit \dBm {dBm}
\DeclareSIUnit \dBW {dBW}
\DeclareSIUnit \dBi {dBi}
\DeclareSIUnit \bpcu {bpcu}

\DeclareFontFamily{U}{mathx}{\hyphenchar\font45}
\DeclareFontShape{U}{mathx}{m}{n}{
      <5> <6> <7> <8> <9> <10>
      <10.95> <12> <14.4> <17.28> <20.74> <24.88>
      mathx10
      }{}
\DeclareSymbolFont{mathx}{U}{mathx}{m}{n}
\DeclareMathSymbol{\bigtimes}{1}{mathx}{"91}

\usepackage{pgfkeys}
    \def\addlegendimage{\csname pgfplots@addlegendimage\endcsname}

\newtheorem{theorem}{Theorem}
\newtheorem{lemma}{Lemma}
\newtheorem{corollary}{Corollary}

\pgfplotscreateplotcyclelist{default}{%
	blue,mark=*\\%
	red,mark=star\\%
	teal,mark=square*\\%
	brown!60!black,mark=otimes*\\%
}

\definecolor{plot1}{RGB}{228,26,28}
\definecolor{plot2}{RGB}{55,126,184}
\definecolor{plot3}{RGB}{77,175,74}
\definecolor{plot4}{RGB}{152,78,163}
\definecolor{plot5}{RGB}{255,127,0}
\definecolor{plot6}{RGB}{166,86,40}
\tikzstyle{fedsatschedule}=[plot1]
\tikzstyle{fedsat}=[plot2]
\tikzstyle{fedisl}=[plot3]
\tikzstyle{fedavg}=[plot4]
\tikzstyle{fedasync1}=[plot5]
\tikzstyle{fedasync2}=[plot6]

\newcolumntype{P}[1]{>{\centering\arraybackslash}p{#1}}

\ifCLASSOPTIONdraftcls
\AtBeginEnvironment{figure}{}
\fi

\AtBeginEnvironment{algorithmic}{\footnotesize}

\usetikzlibrary{patterns}

\begin{document}
\bstctlcite{IEEEexample:BSTcontrol}
\title{Energy-Aware Federated Learning in\\ Satellite Constellations}

{\author{\IEEEauthorblockN{Nasrin Razmi\IEEEauthorrefmark{1}\IEEEauthorrefmark{2}, Bho Matthiesen\IEEEauthorrefmark{1}\IEEEauthorrefmark{2}, Armin Dekorsy\IEEEauthorrefmark{1}\IEEEauthorrefmark{2}, and Petar Popovski\IEEEauthorrefmark{3}\IEEEauthorrefmark{1}}

\IEEEauthorblockA{\IEEEauthorrefmark{1} Dept.\ of Communications Engineering, University of Bremen, Germany}

\IEEEauthorblockA{\IEEEauthorrefmark{2}Gauss-Olbers Space Technology
Transfer Center, University of Bremen, Germany}

\IEEEauthorblockA{\IEEEauthorrefmark{3} Dept. of Electronic Systems, Aalborg University, Denmark}

\IEEEauthorblockA{Emails:  \{razmi, matthiesen, dekorsy\}@ant.uni-bremen.de, petarp@es.aau.dk}

\thanks{This work was funded in part
by the German Research Foundation (DFG) under Germany's Excellence Strategy (EXC 2077 at University of Bremen, University Allowance).}
}}

\maketitle

\begin{abstract}
	\Gls{fl} in satellite constellations, where the satellites collaboratively train a \gls{ml} model, is a promising technology towards enabling globally connected intelligence and the integration of space networks into terrestrial mobile networks. The energy required for this computationally intensive task is provided either by solar panels or by an internal battery if the satellite is in Earth's shadow. Careful management of this battery and system's available energy resources is not only necessary for reliable satellite operation, but also to avoid premature battery aging. We propose a novel energy-aware computation time scheduler for satellite FL, which aims to minimize battery usage without any impact on the convergence speed. Numerical results indicate an increase of more than $3\times$ in battery lifetime can be achieved over energy-agnostic task scheduling. 
\end{abstract}

\glsresetall


\section{Introduction}
Driven by the desire for ubiquitous global connectivity, satellite communications and the space industry are undergoing a major transformation \cite{sweeting2018modern}. 
Especially the paradigm shift towards mega-constellations of interconnected small satellites in \cglspl{leo} led towards space networks nowadays being regarded as a major component of 6G (and beyond) networks \cite{9217520,Abdelsadek2023}. An integral component, both in managing \cite{fontanesi2023artificial} and utilizing \cite{Izzo2022} these constellations, is distributed \cgls{ml}. Among the architectural approaches towards integrating \cgls{ml} into space networks \cite{Chen2022}, \cgls{sfl} \cite{matthiesen2022federated}, where the \cgls{ml} training is performed directly on each satellite, appears to be most promising strategy in the long run. A closely related use case is orbital edge computing \cite{Wu2023a}, where terrestrial devices offload computing tasks, including \cgls{ml} training \cite{Wang2020a}, to satellites.

A considerable challenge in implementing computationally intensive processes in satellite systems is their energy consumption. While, theoretically, solar energy is available in abundance in near-Earth space, the amount of energy a satellite can harvest is practically limited by the size of its solar panels. Moreover, the nature of orbital mechanics places satellites regularly in Earth's shadow, necessitating the usage of batteries to bridge these outages in solar energy. These aspects lead to a limited availability of energy within the satellite. Since this energy is not only used for (secondary) computational tasks but also to supply critical satellite subsystems, careful energy management and scheduling of computing time is strictly necessary to ensure reliable satellite operations \cite{Alagoz2011EnergyEfficiency}. Furthermore, the
lifetime of batteries is limited
by the number of charge/discharge cycles they can withstand. Since replacing a failing battery in a deployed satellite is infeasible, inadequate battery management can have detrimental effect on the lifetime of a satellite system \cite{Chen2022DynamicRouting,Liu2021DRL-ER}. This is not only undesirable from an economical perspective, but also environmentally unsustainable.

To this end, we consider energy-aware task scheduling for \cgls{sfl} with the goal of maximizing the battery lifetime.
\cGls{sfl} is the application of the \cgls{fl} paradigm \cite{mcmahan2017communication} in satellite constellations. It was first studied in \cite{WCL_fedsat} for ground-assisted \cgls{fl}, where the limited connectivity between satellites and \cgls{gs} was identified as the main impairment towards reasonable training performance. Subsequent works focus primarily on improving the algorithm in \cite{WCL_fedsat}, e.g., \cite{Zhou2024AsyncFL, razmi2022scheduling, Yang2024, Zhai2024, Lin2024b}, and enhancing connectivity \cite{razmi2024TCOM,Sch2023Razmi, Elmahallawy2024, Shi2024}.
While energy aspects of \cgls{sfl} are considered in \cite{ConvTime2024Yan, DSFL2022Wu, Han2024CoopFL}, none of these works considers long-term effects of \cgls{sfl} on the battery lifetime. Our proposed algorithm leverages the predictability of sunlight periods and satellite operations to schedule computation time for \cgls{sfl} with respect to the energy demands of more critical satellite subsystems, while using any remaining degrees of freedom to minimize strain on the battery to extend its lifetime.
While we focus primarily on \cgls{sfl}, the proposed algorithm should also be applicable to a wide range of related orbital edge computing scenarios.

\section{System Model}
\subsection{Constellation Configuration}
We consider a constellation with $Q$ orbital planes, where each orbit $q \in \{1, \cdots, Q\}$ contains $K_q$ satellites. The set of all satellites in the $Q$ orbital planes is denoted with $\mathcal{K}$, with a total number of satellites $K=\sum_{q=1}^{Q} K_q$. 
Any satellite $k$ completes an orbit around Earth within a period of $\iota_k = 2 \pi \sqrt{\frac{o_k^3}{\mu}}$, where $o_k$ is the semi-major axis and $\mu = 3.98 \times 10^{14}\,\si{\meter^3/\second^2}$ is the geocentric gravitational constant. For circular orbital planes, semi-major axis is $o_k = r_E + \nu_k$, where $r_E = 6371\,\si{\km}$ and $\nu_k$ are the Earth radius and satellite's altitude above the Earth's surface, respectively. 

\subsection{Computation Model}

Each satellite $k$ gathers dataset $\mathcal D_k$ using its on-board instruments and trains an \cgls{ml} model. The overall goal of satellites is to
collaboratively solve an optimization problem
\begin{equation}
	\min_{\vec w\in\mathds R^d} \frac{1}{D} \sum_{\vec z \in \mathcal D} g(\vec z; \vec w)
	= \min_{\vec w\in\mathds R^d} \sum_{k\in\mathcal K}  \frac{D_k}{D} \sum_{\vec z \in \mathcal D_k} \frac{1}{D_k} g(\vec z; \vec w) \label{eq: opt}
\end{equation}
and learn the global model parameters $\vec w$. In \cref{eq: opt}, $\mathcal D = \bigcup_{k\in\mathcal K} \mathcal D_k$ is the set of data samples of all satellites with the size of $D = \sum_{k\in\mathcal K} D_k$, where $D_k = |\mathcal D_k|$ is the size of dataset of satellite $k$. Moreover, $g(\vec z; \vec w)$ is defined as the loss for a data sample $\vec z\in\mathcal D$ and model parameters $\vec w$. The problem \eqref{eq: opt} is orchestrated by one or several \cglspl{ps} and is solved in overall $N$ iterations without sharing datasets between satellites. 

In iteration $n$, the satellite $k$ performs $M$ local epochs of mini-batch
\cgls{sgd} as \cref{alg:Satellite SGD Procedure} to minimize the local loss function 
$\frac{1}{D_k} \sum\nolimits_{\vec{z}\in\mathcal D_k} h(\vec{z}, \vec{w})$ \cite{razmi2024TCOM}. It finally derives the local model parameters $\vec{w}^{n,M}_{k}$ and transmits it to the \cgls{ps}.
\begin{algorithm}[t!]
	\caption{Satellite Learning Procedure} \label{alg:Satellite SGD Procedure}
	\begin{algorithmic}[1]
		\Procedure{SatLearnProc}{$\vec{w}^{n}$}
			\State \textbf{initialize} $\vec{w}_k^{n,0} = \vec{w}^{n}, \quad m = 0$, \quad learning rate $\eta$ 
        \label{alg:ssgd:init}
			\For {$M$ epochs} \label{alg:ssgd:batchstart}
				\Comment $M$ epochs of mini-batch \cgls{sgd}
				\State $\tilde{\mathcal D}_k \gets $ Randomly shuffle $\mathcal D_k$
				\State $\mathscr B \gets $ Partition $\tilde{\mathcal D}_k$ into mini-batches of size $B$
				\For {each batch $\mathcal B\in\mathscr B$}
				\State $\vec{w}_k^{n,m} \gets \vec{w}_k^{n,m} - \frac{\eta}{|\mathcal B|} \nabla_{\vec{w}} \left(\sum_{\vec z\in\mathcal B} g(\vec z, \vec w) \right)$				
				\EndFor
                        \State $m \gets m + 1$
			\EndFor \label{alg:ssgd:batchend}
		      \label{alg:ssgd:compress}
			\State \Return $D_k \vec{w}^{n,M}_{k}$ \label{alg:ssgd:ret}
		\EndProcedure
	\end{algorithmic}
\end{algorithm}

\subsection{Satellite Energy Management}

\subsubsection{Sunlight and eclipse pattern}

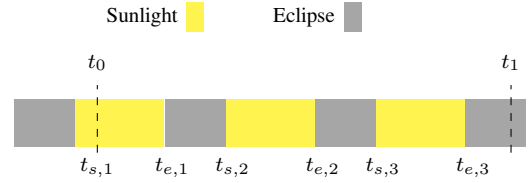
\begin{figure}[h] 
\centering
\begin{tikzpicture}

\node[minimum width=8 mm,text depth=2.5ex,draw, fill=gray!70,draw=none] (0) at (0,0)
 {};
\node[minimum width=11.7mm,text depth=2.5ex,draw, fill=yellow!80,draw=none] (1) at (1,0)
 {};

 \foreach  \X in {3,5}
 {\node[minimum width=16.1mm,text depth=2.5ex,draw, fill=yellow!80,draw=none] (\X) at (1*\X,0)
 {};}
  \foreach  \X in {2,4,6}
  {\node[minimum width=8.1mm,text depth=2.5ex,draw, fill=gray!70,draw=none] (\X) at (1*\X,0)
 {};}

 \draw[dashed]  (0.7,-0.4) -- (0.7,0.5) ;
\draw[dashed]  (6.2,-0.4) -- (6.2,0.5);
\node[] at (0.7,0.8) {\footnotesize{$t_0$}};
\node[] at (6.2,0.8) {\footnotesize{$t_1$}};
\node[] at (0.7,-0.6) {\footnotesize{$t_{s,1}$}};
\node[] at (1.7,-0.6) {\footnotesize{$t_{e,1}$}};
\node[] at (2.5,-0.6) {\footnotesize{$t_{s,2}$}};
\node[] at (3.7,-0.6) {\footnotesize{$t_{e,2}$}};
\node[] at (4.5,-0.6) {\footnotesize{$t_{s,3}$}};
\node[] at (5.7,-0.6) {\footnotesize{$t_{e,3}$}};
\node[minimum width=2 mm,text depth=1ex,draw, fill=yellow!80,draw=none] (2) at (2,1.4)
 {};
 \node[] at (1.3,1.4) {\footnotesize{Sunlight}};
\node[minimum width=2 mm,text depth=1ex,draw, fill=gray!70,draw=none] (2) at (4.1,1.4)
 {};
 \node[] at (3.45,1.4) {\footnotesize{Eclipse}};
 
\end{tikzpicture}
\caption{Satellite's sunlight and eclipse pattern }
\label{figure:sunlight11-eclipse11}
\end{figure}

Satellites are equipped with solar panels to harvest the Sun's radiant energy for power generation in sunlight periods. After each sunlight period, the satellite experiences periods of eclipse, as denoted in \cref{figure:sunlight11-eclipse11}. In an eclipse period, the satellite relies on the energy stored in its battery, usually a Lithium-ion battery, which is charged during sunlight periods. For any designated duration from $t_0$ to $t_1$, we denote the start of each sunlight period and its subsequent eclipse as $t_{s,j}$ and $t_{e,j}$, respectively, where $j \in \mathcal{J} = \{1,\cdots, J\}$ represents the indices of these periods as depicted in \cref{figure:sunlight11-eclipse11}. 

\subsubsection{Energy Consumption}

Satellites utilize the harvested or battery energy to perform various tasks. We categorize the tasks into two groups: non-training, which include all tasks except \cgls{ml} training, and training which refers only to the \cgls{ml} training task. The energy demand of the non-training tasks is determined a~priori by a different subsystem and provided as parameters $E^d_{s,j}$ and $E^d_{e,j}$ to the \cgls{sfl} task scheduler, where $E^d_{s,j}$ and $E^d_{e,j}$ are the energy demands of non-training tasks during the intervals $[t_{s,j}, t_{e,j}]$ and $[t_{e,j}, t_{s,j+1}]$, i.e., the $j$th sunlight and eclipse periods, respectively.
For the training task, the satellites need to train an \cgls{ml} model for a duration of $T_c$, which results in consuming $P_c T_c$ amount of energy, where $P_c$ is the required power for training. We assume that the required training time $T_c$ is smaller than the available time $t_1-t_0$.
Then, the total energy consumption of the satellite is $E^d_{s,j} + \tau_{s,j} P_c$ during sunlight periods, and $E^d_{e,j} + \tau_{e,j} P_c$ during eclipse periods, where $\tau_{s,j}$ and $\tau_{e,j}$ are the computation times scheduled for \cgls{ml} training during these intervals. Clearly, $0 \leq \tau_{s,j} \leq t_{e,j}-t_{s,j}$ and $0 \leq \tau_{e,j} \leq t_{s,j+1}-t_{e,j}$. Further,
to be able to complete \cgls{ml} training, we require 
\begin{equation}
    \sum\limits_{\substack{j=1}}^{J} \tau_{e,j} + \tau_{s,j} = T_c.
    \label{eq: cond_time_training}
\end{equation}

\subsubsection{Satellite battery}
Let $b(t) \in [0, B_\mathrm{max}]$ be the battery's charge level at time instant $t$, where $B_\mathrm{max}$ is the battery's capacity. During eclipse periods, the satellite's energy demand is met solely from the battery. Thus, the remaining charge $b_{s,j+1} = b(t_{s,j+1})$ at the end of eclipse period $j$ is
\begin{equation} \label{battery_level_sunlight_period}
	b_{s,j+1} = b_{e,j} - E^d_{e,j}- P_c \tau_{e,j},
\end{equation}
where $b_{e,j} = b(t_{e,j})$ is the battery level at the beginning of eclipse period $j$. During sunlight periods, the battery is recharged from the solar panels. Provided the harvested energy during sunlight period $j$, i.e., in the time interval $[t_{s,j}, t_{e,j}]$ is $E^h_{s,j}$, the energy available for charging the battery is $E^h_{s,j} - E^d_{s,j} - P_c\tau_{s,j} $. Thus, the battery level at the begin of the following eclipse period is
\begin{equation}
         b_{e,j} = \min \{b_{s,j} + E^h_{s,j} - E^d_{s,j} - P_c \tau_{s,j},\  B_{\max}\}.
\end{equation}
This implies that excess energy goes to waste once the battery is fully charged. Since the battery should not be discharged further during sunlight periods, we require
\begin{equation}
	E^h_{s,j} - E^d_{s,j} - P_c \tau_{s,j}  \ge 0,
\end{equation}
which implies $E^h_{s,j} \ge E^d_{s,j} + P_c \tau_{s,j}$.

\subsubsection{Battery aging}\label{sec:aging}

To extend the operational life of the satellites, it is crucial to prolong the lifetime of the batteries. An effective parameter to evaluate the battery's lifetime is cycle life, defined as the number of charge and discharge cycles a battery can complete before its performance degradation \cite{qadrdan2018smart}. With each charge/discharge cycle, the battery degrades. The battery last longer with a smaller discharge level \cite{Bu-808}.

The proportion of energy extracted from the battery within an eclipse duration relative to its full capacity is defined as \cgls{dod}, expressed as $d(t) = \frac{B_{\max}-b(t)}{B_{\max}}$ at any given time $t$. In this paper, we focus on the effect of \cgls{dod} on cycle life \cite{Yang2016EnergyRouting}.
A higher \cgls{dod} results in a shorter battery lifetime, meaning a higher consumption of cycle life. 
The cycle life consumption for Lithium-ion batteries is defined as
\begin{align} \label{eq: dod_definition1}
      l(t_1,t_2) = &\int_{d(t_1)}^{d(t_2)} 10^{a(d-1)} \left(1+a \ln10 \cdot d \right) \text{d}d \\ 
          = &10^{a(d(t_2)-1)}d(t_2) - 10^{a(d(t_1)-1)}d(t_1) \nonumber,     \end{align}
if $d(t_2) > d(t_1)$; otherwise, it is 0, and $a$ is a positive constant which is determined by the battery specifications \cite{Yang2016EnergyRouting}.

\section{Cycle life Minimization}
We aim to schedule the computation time for \cgls{ml} training such that all non-training energy demands are met and the cost on the batteries life is minimized. Since the battery is never discharged during sunlight periods, the total cost on the battery lifetime during $[t_0, t_1]$ is, following \cref{sec:aging},
\vspace{-0.1 cm}
\begin{equation} \label{eq: cycle_life_eclipse_wo_simplify}
	\sum_{j=1}^{J} 10^{a(\hat d_{e,j}-1)}\hat d_{e,j} - 10^{a(d_{e,j}-1)}d_{e,j},
\end{equation}
where $d_{e,j} = d(t_{e,j})$ and $\hat d_{e,j} = d(t_{s,j+1})$ are the \cgls{dod} at the beginning and end of the $j$th eclipse period $[t_{e,j}, t_{s,j+1}]$, respectively. Thus, the optimal \cgls{ml} training schedule with respect to the battery lifetime is the solution to
\begin{subequations}
	\label{eq:WS1a}
	\begin{align}
		\min_{\substack{\forall j: \tau_{s,j}, \tau_{e,j}, b_{e,j},\\ b_{s,j+1}, d_{e,j}, \hat{d}_{e,j}}}\quad
		& \sum_{j=1}^{J} 10^{a(\hat{d}_{e,j}-1)}\hat{d}_{e,j} - 10^{a(d_{e,j}-1)}d_{e,j} \label{eq: opt_cost_func}\\
	\mbox{s.t.}\quad
        & b_{s,j+1} = b_{e,j} - E^d_{e,j}- P_c \tau_{e,j}, \forall \in \mathcal J \label{eq: battery_sunlight}\\
        \begin{split}
        & b_{e,j} =\min\{b_{s,j}+E^h_{s,j}-\\&\hspace{2.3em}E^d_{s,j}-P_c \tau_{s,j},\ B_{\max}\},\ \forall j \in \mathcal J, 
        \end{split}\label{eq: battery_start_eclipse}\\
		& E^h_{s,j} - E^d_{s,j} - P_c \tau_{s,j}  \ge 0,  \ \forall j \in \mathcal J, \label{eq: cond_discharge_sunlight} \\
	  & d_{e,j}=\frac{B_{\max}-b_{e,j}}{B_{\max}}, \ \forall j \in \mathcal J, \label{eq: dod_start_eclipse} \\
       &\hat{d}_{e,j} = \frac{B_{\max}-b_{s,j+1}}{B_{\max}}, \ \forall j \in \mathcal J, \label{eq: dod_end_eclipse}  \\
	& \sum\limits_{\substack{j=1}}^{J}  \tau_{s,j} + \tau_{e,j}  =  T_c,\label{eq: training_period1} \\
        & 0 \leq \tau_{s,j}  \leq t_{e,j}- t_{s,j},\  \forall j \in \mathcal J, \label{eq: cons_training_time_sunlight}\\
        & 0 \leq \tau_{e,j}  \leq t_{s,j+1}- t_{e,j},\  \forall j \in \mathcal J, \label{eq: cons_training_time_eclipse}\\
		& 0 \le b_{s,j+1} \le B_\mathrm{max},\  \forall j \in \mathcal J, \label{eq: cons_battery_sunlight}\\
		& 0 \le b_{e,j} \le B_\mathrm{max},\  \forall j \in \mathcal J, \label{eq: cons_battery_eclipse} 
	\end{align}
\end{subequations}
with $b_{s,1} = B_{0}$, where $B_0$ is the initial battery charge at $t_0$. Conditions \cref{eq: battery_sunlight} and 
\cref{eq: battery_start_eclipse} signify the battery level at the beginning of the sunlight and eclipse periods, respectively. Additionally, \cref{eq: cond_discharge_sunlight} implies that the battery should not be discharged during sunlight periods. \cgls{dod} at the beginning and end of the eclipse periods are defined in \cref{eq: dod_start_eclipse} and \cref{eq: dod_end_eclipse}. The training periods during sunlight and eclipse should meet
\cref{eq: training_period1}, with constraints defined as \cref{eq: cons_training_time_sunlight} and \cref{eq: cons_training_time_eclipse} for these periods, respectively. Moreover, \cref{eq: cons_battery_sunlight} and \cref{eq: cons_battery_eclipse} indicate that the battery level must remain above 0 and below the battery's capacity.
\vspace{-0.2 cm}
\subsection{Equivalent Problem}
Problem~\cref{eq:WS1a} is nonconvex due to \cref{eq: opt_cost_func,eq: battery_start_eclipse}. However, we can relax \cref{eq: battery_start_eclipse} to obtain
\begin{subequations}
	\label{eq:WS2}
	\begin{align}
		\min_{\substack{\forall j: \tau_{s,j}, \tau_{e,j}, b_{e,j},\\ b_{s,j+1}, d_{e,j}, \hat{d}_{e,j}}}\quad 
		& \sum_{j=1}^{J} 10^{a(\hat{d}_{e,j}-1)}\hat{d}_{e,j} - 10^{a(d_{e,j}-1)}d_{e,j} \label{eq: opt_cost_func_updated}\\
	\mbox{s.t.}\quad
    \begin{split}
        & b_{e,j} \le \min\{b_{s,j} + E^h_{s,j} -\\&\hspace{2.3em}  E^d_{s,j} - P_c \tau_{s,j},\ B_{\max}\},\ \forall j \in \mathcal J,
    \end{split}\label{c1relax}\\
		& \text{\cref{eq: battery_sunlight,eq: cond_discharge_sunlight,eq: dod_start_eclipse,eq: dod_end_eclipse,eq: training_period1,eq: cons_training_time_sunlight,eq: cons_training_time_eclipse,eq: cons_battery_sunlight,eq: cons_battery_eclipse}},
	\end{align}
\end{subequations}
which has a convex feasible set and is equivalent to \cref{eq:WS1a}.
\begin{lemma}
	Any optimal solution to \cref{eq:WS2} is an optimal solution to \cref{eq:WS1a} (and vice versa).
\end{lemma}
\begin{IEEEproof}
	Consider problem~\cref{eq:WS1a}. For an arbitrary $n\in\mathcal J$, replace \cref{eq: battery_start_eclipse} by \cref{c1relax} and let $\tau_{s,1}^\star, \dots, \hat d_{e,j}^\star$ be an optimal solution to this new problem. Assume this solution satisfies
	\begin{equation}
		b_{e,n}^\star < \min\{b_{s,n}^\star+ E^h_{s,n} - E^d_{s,n} - P_c \tau_{s,n}^\star,\ B_{\max}\}.
	\end{equation}
	Then, we can find a $\tilde b_{e,n} = b_{e,n}^\star + \varepsilon$ with $\varepsilon > 0$ that satisfies \cref{c1relax} and \cref{eq: cons_battery_eclipse}. From \cref{eq: battery_sunlight}, we obtain $\tilde b_{s,n+1} = b_{s,n+1} + \varepsilon$. Further, $\tilde d_{e,n} = d_{e,n}^\star - \delta$ and $\tilde{\hat d}_{e,n} = \hat d_{e,n}^\star - \delta$ with $\delta = \frac{\varepsilon}{B_\mathrm{max}} > 0$. The cycle life consumption for these \cglspl{dod} is
	\begin{align}
		&10^{a(\hat{d}_{e,n}^\star-\delta-1)}(\hat{d}_{e,n}^\star-\delta) - 10^{a(d_{e,n}^\star-\delta-1)}(d_{e,n}^\star-\delta)
		\\{}={}&
		10^{-a\delta}\left( 10^{a(\hat{d}_{e,n}^\star-1)}\hat{d}_{e,n}^\star - 10^{a(d_{e,n}^\star-1)}d_{e,n}^\star \right) \notag
		\\&
		\qquad\qquad- \delta \left(10^{a(\hat{d}_{e,n}^\star-\delta-1)} - 10^{a(d_{e,n}^\star-\delta-1)}\right)
		\\{}\le{}&
		10^{-a\delta}\left( 10^{a(\hat{d}_{e,n}^\star-1)}\hat{d}_{e,n}^\star - 10^{a(d_{e,n}^\star-1)}d_{e,n}^\star \right)
		\label{ineq1}
		\\{}\le{}&
		10^{a(\hat{d}_{e,n}^\star-1)}\hat{d}_{e,n}^\star - 10^{a(d_{e,n}^\star-1)}d_{e,n}^\star
		\label{ineq2}
	\end{align}
	where \cref{ineq1} follows from $\hat d_{e,n} \ge d_{e,n}$ and \cref{ineq2} is due to $a, \delta > 0$.
	Finally, for all $j > n$, we also obtain new feasible $\tilde b_{e,j} \ge b_{e,j}^\star$ and $\tilde b_{s,j+1} \ge b_{s,j+1}^\star$ that further decrease the objective value (by the same argument as before). This implies that $\tau_{s,1}^\star, \dots, \hat d_{e,j}^\star$ cannot be an optimal solution to \cref{eq:WS2} and, further, that any optimal solution satisfies \cref{c1relax} with equality.
\end{IEEEproof}

Due to this equivalence, we can obtain a solution to \cref{eq:WS1a} by solving \cref{eq:WS2}. However, \cref{eq:WS2} is still a nonconvex optimization problem due to the objective being a \cgls{dc}. Thus, we cannot solve it directly with conventional convex optimization methods. Instead, we design an iterative algorithm based on the majorization-minimization principle \cite{Sun2017} to find stationary points of \cref{eq:WS2}.

\section{First-order Optimal Solution Algorithm}
Problem \cref{eq:WS2} falls in the class of \cgls{dc} programming problems with convex feasible set, i.e., a continuous optimization problem
\vspace{-0.3 cm}
\begin{equation} \label{opt:dc}
	\min_{\vec x} f(\vec x) = u(\vec x) - v(\vec x) \quad\mathrm{s.t.}\quad \vec x \in \mathcal X,
\end{equation}
where $\mathcal X$ is a closed convex subset of $\mathds R^n$ and $u, v$ are continuously differentiable convex functions on $\mathcal X$. The concave-convex procedure \cite{Yuille2003,Sriperumbudur2009} obtains stationary points of \cref{opt:dc} by solving a sequence of convex programs
\begin{equation} \label{opt:cccp}
	\vec x^{(l+1)} \in \argmin_{\vec x \in \mathcal X} u(\vec x) - \vec x^T \nabla v(\vec x^{(l)}).
\end{equation}
This is an instance of the more general majorization-minimization framework where the nonconvex part of the objective is linearized. Within this context, the objective of \cref{opt:cccp} is known as the \emph{surrogate function} of $f(\vec x)$.

\subsection{Concave-convex procedure for \cref{eq:WS2}}
For ease of notation, define the vectors $\vec d = (d_{e,1}, \dots, d_{e,J})$, $\hat{\vec d} = (\hat d_{e,1}, \ldots, \hat d_{e,J})$, and $\vec x$ to hold all optimization variables in \cref{eq:WS2}. We identify $\mathcal X$ as $\{\vec x \mid \text{\cref{eq: battery_sunlight}, \cref{c1relax}, \cref{eq: cond_discharge_sunlight,eq: dod_start_eclipse,eq: dod_end_eclipse,eq: training_period1,eq: cons_training_time_sunlight,eq: cons_training_time_eclipse,eq: cons_battery_eclipse,eq: cons_battery_sunlight}}\}$, $u(\vec x) = \sum_{j=1}^{J} 10^{a(\hat{d}_{e,j}-1)}\hat{d}_{e,j}$, and, with a minor abuse of notation, $v(\vec x) = v(\vec d) = \sum_{j=1}^{J} 10^{a(d_{e,j}-1)}d_{e,j}$ in \cref{opt:cccp}. First, observe that this is indeed a \cgls{dc} program since $\mathcal X$ is a convex set (trivial) and $u, v$ are convex functions, as established next.
\begin{lemma} \label{lem:scvx}
	For $a > 0$, the function $\sum_{j = 1}^J y_j 10^{a (y_j - 1)}$ is strictly convex on $\mathds R^J_+$ with respect to $y_1, \dots, y_J$, i.e., for all $y_j \ge 0$, $1 \le j \le J$.
\end{lemma}
\begin{IEEEproof}
	The Hessian of $h = \sum_{j = 1}^J y_j 10^{a (y_j - 1)}$ is a diagonal matrix with the partial derivatives $\partial^2 h / \partial y_j^2$ on the diagonal. Thus, it is strictly positive definite if and only if $\partial^2 h / \partial y_j^2 > 0$ for all $j$. From $\partial^2 h / \partial y_j^2 = a \ln{10} (a y_j \ln{10}+2) 10^{a (y_j-1)}$, we see that this is the case if and only if $a y \ln{10}+2 > 0$ for $a > 0$. Thus, $y_j \ge 0$ is sufficient for strong convexity.
\end{IEEEproof}
\begin{corollary} \label{cor:cvx}
	The functions $u$ and $v$ are convex on $\mathcal X$.
\end{corollary}
\begin{IEEEproof}
	From \cref{eq: dod_end_eclipse,eq: dod_start_eclipse,eq: cons_battery_sunlight,eq: cons_battery_eclipse}, we have $\hat d_{e,j} \ge 0$ and $d_{e,j} \ge 0$ for all $j$. Since $a > 0$ by definition, \cref{lem:scvx} is applicable.
\end{IEEEproof}

Observe that the nonconvexity in \cref{eq:WS2} stems solely from $v(\vec d)$. Thus, we linearize the objective only in $\vec d$ to obtain the following surrogate problem
\begin{equation} \label{opt}
	\vec x^{(l+1)} \in \argmin_{\vec x \in \mathcal X} u(\vec x) - \vec d^T \nabla v(\vec d^{(l)}),
\end{equation}
\vspace{-0.15 cm}
where
\vspace{-0.15 cm}
\begin{equation*}
	\vec d^T \nabla v(\vec d^{(l)}) = \sum_{j=1}^J d_{e,j}  10^{a(\tilde d_{e,j}-1)}(1 + a \tilde d_{e,j} \ln{10} ).
\end{equation*}
\vspace{-0.05 cm}
This results in the concave-convex procedure stated in \cref{alg}.
After initialization, it solves the convex program in line~\ref{alg:1} for $\vec x^{(l)}$ and sets the next approximation point $\vec x^{(l+1)}$ to an optimal solution of this problem. Note that $\vec d^{(l+1)}$ is implicitly defined as it is part of $\vec x^{(l+1)}$. This is repeated until convergence in $\vec d^{(l)}$ is observed.
Upon termination, the final $\vec d^{(l)}$ will be within an $\varepsilon$-region of the point $\vec d^\star$, where $\vec d^\star$ is such that the corresponding $\vec x^\star$ is a stationary point of \cref{eq:WS2}.
The initial $\vec x^{(0)}$ is set to a feasible point \cref{eq:WS2}, which is easily obtained due to the feasible set $\mathcal X$ being convex.

\begin{algorithm}[t]
	\caption{Concave-convex procedure for \cref{eq:WS2}}
	\label{alg}
	\begin{algorithmic}[1]
		\State\textbf{Initialize} $l = 0$, $\lambda > 0$, $\varepsilon > 0$, and $\vec x^{(0)}$ to some feasible point of \cref{eq:WS2}. \label{alg:2}
		\Repeat
		\State Set $\vec x^{(l+1)}$ to an optimal point of \cref{opt} \label{alg:1}
		\State Update $l \gets l + 1$
		\Until{$\Vert \vec d^{(l)} - \vec d^{(l-1)} \Vert \le \varepsilon$}
	\end{algorithmic}
\end{algorithm}

Convergence of \cref{alg} is established formally below.
Suitable convergence results for \cref{opt:cccp} are derived in Theorems~4 and~8 of \cite{Sriperumbudur2009}. There, Theorem~8 provides much stronger convergence guarantees than Theorem~4 but requires strong convexity of $u$ and $v$ in $\vec x$. This is not the case here, as established in \cref{cor:cvx}. However, the proof of \cite[Thm. 8]{Sriperumbudur2009} can be modified to cover \cref{opt} as follows.
\begin{theorem} \label{thm}
	For all $\varepsilon \ge 0$, the sequence $\{ \vec x^{(l)} \}_{l}$ generated by \cref{alg} converges towards a point $\bar{\vec x}$. For $\varepsilon = 0$, this is a stationary point of \cref{eq:WS2}. Otherwise, \cref{alg} terminates after a finite number of iterations and $\bar{\vec x}$ is within an $\delta$-region of some stationary point of \cref{eq:WS2} for some small $0 < \varepsilon < \delta$, i.e., $\Vert \vec x^\star - \bar{\vec x} \Vert \le \delta$ for some stationary point $\vec x^\star$.
\end{theorem}
\begin{IEEEproof}
	Observe that the feasible set $\mathcal X$ is a finite intersection of affine constraints. By Slater's condition, these constraints are qualified if some feasible point exists \cite[Sec.~5.2.3]{Boyd2004}.
	Further,
	define the point-to-set map $\mathcal A$ (cf.~\cite[Sec.~3]{Sriperumbudur2009}) as the projection of \cref{opt} onto $\vec d$. Let $\vec d^\star$ be a generalized fixed point of $\mathcal A$ and $\vec x^\star$ the corresponding solution of \cref{opt}. Then, by the same argument as in \cite[Lemma~5]{Sriperumbudur2009}, $\vec x^\star$ is a stationary point of \cref{eq:WS2}. Moreover,
	note that the \cgls{rhs} of \cref{opt} is equivalent to
	\begin{equation}
		\argmin_{\vec x \in \mathcal X} u(\vec x) - v(\vec d^{(l)}) - (\vec d - \vec d^{(l)})^T \nabla v(\vec d^{(l)})
	\end{equation}
	Thus, for any $\vec d^{(l)} \neq \vec d^{(l+1)}$,
	\begin{align*}
		\MoveEqLeft f(\vec x^{(l+1)}) = u(\vec x^{(l+1)}) - v(\vec d^{(l+1)}) \\
		&< u(\vec x^{(l+1)}) - v(\vec d^{(l)}) - (\vec d^{(l+1)} - \vec d^{(l)})^T \nabla v(\vec d^{(l)}) \\
		&= g(\vec x^{(l+1)}, \vec x^{(l)})
		\le g(\vec x^{(l)}, \vec x^{(l)}) = f(\vec x^{(l)})
	\end{align*}
	due to $v$ being strongly convex in $\vec d$. This establishes that
	the map $\mathcal A$ is strictly monotonic with respect to $f$.
	Since the feasible set $\mathcal X$ is closed and bounded, i.e., compact, its projection onto $\vec d$ is as well and, with \cite[Remark~7]{Sriperumbudur2009}, $\mathcal A$ is uniformly compact and closed. Hence, by virtue of \cite[Thm. 3]{Sriperumbudur2009}, all limit points of $\{ \vec d^{(l)} \}_l$ are fixed points of $\mathcal A$ and $\Vert \vec d^{(l)} - \vec d^{(l-1)} \Vert \to 0$. Further, since $v$ is a strictly increasing function in $\vec d$, the set of fixed points of $\mathcal A$ is finite and $\{ \vec d^{(l)} \}_l$ converges to a single limit point $\vec d^\star$. The associated point $\vec x^\star$ is a stationary point of \cref{eq:WS2} (see argument above).
	Finally, since the sequence $\{ f(\vec d^{(l)}) \}_l$ is strictly decreasing and $\Vert \vec d^{(l)} - \vec d^{(l-1)}\Vert$ converges continuously to 0, there exists some $L < \infty$ such that $\Vert \vec d^{(l)} - \vec d^{(l-1)}\Vert < \varepsilon$ for all $l > L$ if $\varepsilon > 0$. It is trivial to show that $\Vert \vec d^{(l)} - \vec d^{(l-1)}\Vert < \varepsilon$ implies $\Vert \vec x^{(l)} - \vec x^{(l-1)}\Vert < \delta$ for some $\delta > \varepsilon$.
\end{IEEEproof}

\Cref{alg} does not guarantee that the stationary point is a local minimum of \cref{eq:WS2}. However, unless $\vec x^{(0)}$ is stationary, which would result in \cref{alg} terminating after the first iteration, the obtained point cannot be a local maximum due to $\{ u(\vec x^{(l)}) - v(\vec d^{(l)}) \}_l$ being strictly decreasing.

Convergence of \cref{alg} can be strengthened to $\bar{\vec x}$ being within an $\varepsilon$-region of a stationary point of \cref{eq:WS2} by replacing the termination criterion with $\Vert \vec x^{(l)} - \vec x^{(l-1)} \Vert \le \varepsilon$. This, however, requires regularization in \cref{opt} with $\lambda \Vert \vec x - \vec x^{(l)} \Vert^2$ for some small $\lambda > 0$ to ensure convergence of $\{ \vec x^{(l)} \}_l$ to a single limit point.

\section{PERFORMANCE EVALUATION} \label{sec:numeval}
We evaluate the performance of the proposed energy-aware \cgls{fl}
approach by considering \num{20} satellites from the Starlink constellation \cite{Starlink}. Supported by two \cglspl{gs}, one located in Germany and the other in Japan, these satellites participate in a synchronous \cgls{fl} process for \num{96} hours \cite{WCL_fedsat}. 
We partition this period into $N$ equal time-slots, each equivalent to an iteration of \cgls{fl}.

At the end of the $n$th time-slot, one of the \cglspl{gs}, which is chosen as the \cgls{ps}, updates the global model parameters as $\vec{w}^{n+1} = \sum_{k=1}^{K} \alpha_k^n \frac{D_k}{D} \vec{w}_{k}^{n,M}$, where the local model parameters of the $k$th satellite, $\vec{w}_{k}^{n,M}$, is received
 either directly or through the other \cgls{gs}, and $\alpha_k^n=1$ if the satellite $k$ participates in the $n$th iteration, otherwise $\alpha_k^n=0$. 
During each time-slot, only those satellites can participate in \cgls{fl} which are capable of receiving the global model parameters, training the \cgls{ml} model for a period of $T_c$, and returning their updated local model parameters to the \cgls{ps}. Note that to communicate the model parameters, the satellite should be visible to one of the \cglspl{gs}. Afterwards, the \cgls{ps} sends back the updated global model parameter $\vec{w}^{n+1}$ to the satellites for the next iteration.
Upon receiving the updated parameters, each participating satellite schedules its training, taking into account the predictability of both visibility to \cglspl{gs} and sunlight/eclipse periods.

We compare \cgls{dod} and consumed cycle life for the battery of the satellites using our proposed approach, which we call energy-aware, with those of the state-of-the-art one \cite{WCL_fedsat}, which we call energy-agnostic. In the energy-aware approach, in each time-slot, the participating satellites solve \cref{eq:WS2} to decide how much time they should assign in each sunlight or eclipse for training the \cgls{ml} model. However, in the energy-agnostic one, the participants start the training after receiving the model parameters without considering being in sunlight or eclipse.
\begin{figure}
\centering
    \begin{tikzpicture}
\begin{axis}[
    ybar,
    enlargelimits=0.075,
    legend style={at={(0.5,-0.2)},
    legend pos=north east,legend columns=2},
    bar width=0.1 cm,
    xlabel={Index of time-slot},
    ylabel={Depth of discharge (DoD)},
    symbolic x coords={8,9,21, 22,23,33,41,46},
    xtick=data,
    ymajorgrids=true,
    grid style=dashed,
    ymin=0,
    ymax=1.1,
    width=\axisdefaultwidth,
    height=.89*\axisdefaultheight,
    ]
\addplot[draw=green, pattern color = green, pattern = north west lines] coordinates {(8, 0.015) (9, 0.015) (21, 0.4145833403952306) (22, 0.015) (23, 0.015) (33, 0.015) (41, 0.015) (46, 0.015)};

\addplot[draw=red, pattern color = red, pattern = north west lines] coordinates {(8, 0.5) (9, 0.385) (21, 0.41458333333333336) (22, 0.43833333333333335) (23, 0.5) (33, 0.5) (41, 0.2175) (46, 0.35291666666666666)};

\addplot[fill=green!60] coordinates {(8, 0.249) (9, 0.249) (22, 0.21916667604094844) (23, 0.2499791686309282) (33, 0.23747918269368837) (41, 0.1087500006604756) (46, 0.17645836111947907)};

\addplot[fill=red!60] coordinates {(8, 0.5208333333333334) (9, 0.25) (22, 0.875) (23, 0.7715000000000024) (33, 0.5458333333333333) (41, 0.875) (46, 0.875)};

\legend{\footnotesize{E-Awa., $T_c=20 \ \text{min}$}, \footnotesize{E-Agn., $ T_c=20 \ \text{min}$},\footnotesize{E-Awa., $T_c=80 \ \text{min}$}, \footnotesize{E-Agn., $ T_c=80 \ \text{min}$}}
\end{axis}
\end{tikzpicture}
\caption{Depth of discharge (DoD) for the battery of satellite $k=2$ with respect to indexes of time-slot. E-Agn. and E-Awa. stand respectively for energy-agnostic and energy-aware FL algorithms and $T_c$ denotes required time for training the \cgls{ml} model.}
\label{fig:DoD}
\end{figure}
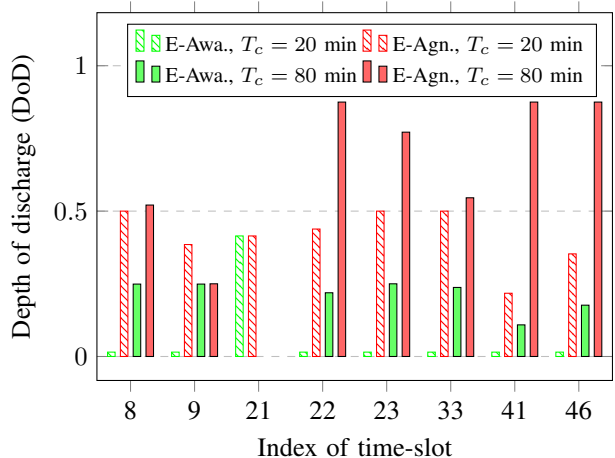
With $N=50$ time-slots and two training duration of $T_c= \SI{20}{\minute}$ and $T_c=\SI{80}{\minute}$, \cref{fig:DoD} shows \cgls{dod} for the satellite $k=2$ during the time-slots in which it can participate in \cgls{fl}. The power consumption for training the \cgls{ml} model is set to $\SI{50}{\watt}$, the battery capacity to $\SI{2000}{\watt\minute}$, and the battery specification parameter $a$ to \num{0.8} as \cite{Yang2016EnergyRouting}. We assume the satellites consume energy solely for training the \cgls{ml} model. Additionally, the harvested energy during sunlight suffices to charge fully the battery. As we see in \cref{fig:DoD}, the satellite $k=2$ participates in \cgls{fl} process only in the time-slots \{8, 9, 21, 22, 23, 33, 41, 46\} when $T_c=\SI{20}{\minute}$, and the same time-slots except \num{21} when $T_c=\SI{80}{\minute}$ due to its visibility pattern. During time-slot \num{21}, there is not sufficient time, from the instant that the satellite receives the global model parameters to the latest possible time it can return the updated local model to the \cgls{ps}, to accommodate the required training duration of $T_c=\SI{80}{\minute}$.
 
As we see in \cref{fig:DoD}, the energy-aware approach leads to a notably lower \cgls{dod} compared to the energy-agnostic approach. Specifically, except for time-slot 21, the \cgls{dod} remains at \num{0} for all other time-slots in the energy-aware approach. This is because the satellite prioritizes utilizing sunlight periods for training without drawing energy from its battery. Only when sunlight periods are insufficient, the satellite starts to use a portion of the eclipse duration for the remaining training which draws energy from its battery, as observed in the time-slot \num{21}. Moreover, the algorithm strives to evenly distribute the remaining training time across all eclipse periods to minimize the impact on cycle life.
However, with the energy-agnostic approach, \cgls{dod} is significantly higher since the satellite trains the model regardless of the current energy source.

\begin{figure}
\begin{tikzpicture}
\begin{axis}[
yminorgrids = true, 
legend entries = { \small{E-Agn.},\small{E-Awa.}},
xlabel={\small{Time [h]}},
xlabel={\centering{\small{Battery capacity [$\SI{}{\watt\minute}$]}}},
ylabel={Consumed cycle life},
grid=major,
grid=minor,
grid = major,
width=1*\axisdefaultwidth,
height=0.7*\axisdefaultheight,
legend cell align=left,
legend pos=north east,
]

\addplot[color=black!20!red,,mark=triangle*, mark options={scale=2,fill=white},line width=1.1pt] table [x=battery_capacity, y=plot_lifecyle_battrey_capacity_baseline, col sep=comma] {images/cycle_life_updated.csv};

\addplot[color=black!40!green,,mark=square*, mark options={scale=1.4,fill=white},line width=1.1pt] table [x=battery_capacity, y=plot_lifecyle_battrey_capacity_proposed, col sep=comma] {images/cycle_life_updated.csv};

\end{axis}
\end{tikzpicture}
\caption{Consumed cycle life with respect to battery capacity during 96 hours, considering required time for training of $T_c = \SI{80}{\min}$. E-Agn. and E-Awa. stand respectively for energy-agnostic and energy-aware FL algorithms.}
\label{fig: cycle_life_battery_capacity}
\end{figure}
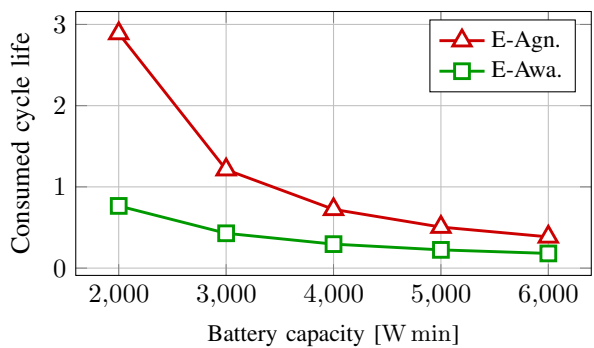
 
\cref{fig: cycle_life_battery_capacity} shows the consumed cycle life of the battery of satellites on average over the total 96 hours with respect to the battery capacity while $T_c = \SI{80}{\min}$.
As we see, the energy-aware approach makes satellites consume lower cycle life of their battery. Specifically, if we consider a battery with capacity of $\SI{2000}{\watt\minute}$ for satellites, by energy-agnostic algorithm, \num{2.88} cycles are consumed, whereas by the energy-aware algorithm, only \num{0.76} cycles are used, meaning over three-fold more cycle life consumption with energy-agnostic. Considering a battery with a total cycle life of \num{800} \cite{Bu-808}, if the satellite employs the energy-agnostic approach, it can operate only for approximately \num{3} years. However, by the energy-aware approach, the satellite's operational lifetime extends to over \num{11} years.
\vspace{-0.15 cm}
\section{Conclusions} \label{sec: conc}

Satellites use solar energy during sunlight periods but depend on their batteries during eclipses. However, frequent use of batteries decreases their lifetime. To enhance the lifetime of satellite batteries, in this paper, we formulated an optimization problem. The aim is to schedule on-board FL training tasks such that the use of the satellite's batteries is reduced. We solved the optimization problem with successive convex approximation. Our numerical results show that the introduced approach increases the lifetime of batteries significantly.
\balance
\bibliography{IEEEtrancfg,IEEEabrv,references}

\end{document}